\documentclass[letterpaper, 10pt, conference]{ieeeconf} 
\IEEEoverridecommandlockouts  
\usepackage[dvipdfmx]{graphicx}

\usepackage{indentfirst}
\usepackage{array,multirow}
\usepackage{fancyhdr}
\usepackage{graphicx}
\usepackage{float}
\usepackage{booktabs}
\usepackage{multirow}
\usepackage{here}
\usepackage{comment}
\usepackage[switch]{lineno,xcolor}

\usepackage{subfigure}
\usepackage{caption}
\usepackage{siunitx}
\usepackage{hyperref}
\usepackage{cite}
 
\newcommand{\etal}{et~al.\ }
\newcommand{\eg}{e.\,g.,\ }
\newcommand{\ie}{i.\,e.,\ }

\title{\LARGE \bf
Autonomous Vehicles Drive into Shared Spaces:\\eHMI Design Concept Focusing on Vulnerable Road Users
\vspace{-2mm}
}

\author{
Yang Li$^{1,*}$, 
Hao Cheng$^{2,*},$ 
Zhe Zeng$^{3,*}$, 
Hailong Liu$^{4,*}$
Monika Sester$^{2}$
\vspace{-2mm}
\thanks{$^{1}$Yang Li is with the ifab-Institute of Human and Industrial Engineering, Karlsruhe Institute of Technology, Engler-Bunte-Ring 4, Karlsruhe, Germany. {\tt\small yang.li@kit.edu}}%
\thanks{$^{2}$Hao Cheng and Monika Sester are with the Institute of Cartography and Geoinformatics, Leibniz University Hannover, Appelstr. 9a, 30167, Hannover, Germany. 
{\tt\small \{hao.cheng, monika.sester\}@ikg.uni-hannover.de}}%
\thanks{$^{3}$Zhe Zeng is with the Chair of Human-Machine Systems, Technical University of Berlin, Marchstr. 23, 10587, Berlin, Germany. {\tt\small zhe.zeng@mms.tu-berlin.de}}%
\thanks{$^{4}$Hailong Liu is with Graduate School of Informatics, Nagoya University, Furo-cho, Chikusa-ku, Nagoya, Aichi, 464-8601, Japan. {\tt\small lhl881210@live.com}}%
\thanks{$^*$These four co-first authors contribute equally to the work.}
}

\begin{document}
\maketitle
\thispagestyle{empty}
\pagestyle{empty}

\begin{abstract}
In comparison to conventional traffic designs, \textit{shared spaces} promote a more pleasant urban environment with slower motorized movement, smoother traffic, and less congestion.
In the foreseeable future, shared spaces will be populated with a mixture of autonomous vehicles~(AVs) and vulnerable road users~(VRUs) like pedestrians and cyclists.
However, a driver-less AV lacks a way to communicate with the VRUs when they have to reach an agreement of a negotiation, which brings new challenges to the safety and smoothness of the traffic.
To find a feasible solution to integrating AVs seamlessly into shared-space traffic, we first identified the possible issues that the shared-space designs have not considered for the role of AVs. 
Then an online questionnaire was used to ask participants about how they would like a driver of the manually driving vehicle to communicate with VRUs in a shared space.
We found that when the driver wanted to give some suggestions to the VRUs in a negotiation, participants thought that the communications via the driver's body behaviors were necessary. 
Besides, when the driver conveyed information about her/his intentions and cautions to the VRUs, participants selected different communication methods with respect to their transport modes (as a driver, pedestrian, or cyclist).
These results suggest that novel eHMIs might be useful for AV-VRU communication when the original drivers are not present.
Hence, a potential eHMI design concept was proposed for different VRUs to meet their various expectations.
In the end, we further discussed the effects of the eHMIs on improving the sociality in shared spaces and the autonomous driving systems.
\end{abstract}

\section{Introduction}
In the 1970s, the concept of shared spaces as a traffic design was introduced by the Dutch traffic engineer Hans Monderman~\cite{emma2006shared}. 
It was later formally defined by Reid~\cite{reid2009dft} as ``a street or place designed to improve pedestrian movement and comfort by reducing the dominance of motor vehicles and enabling all users to share the space rather than follow the clearly defined rules implied by more conventional designs''. 
Shared-space designs largely remove road signs, markings, and traffic lights with no or minimum traffic regulations to allow vehicles with a limited speed to directly interact with pedestrians and cyclists. 
In other words, road users are not separated by time or space segregation according to their transport modes, \eg as a driver, a pedestrian, or a cyclist.
They negotiate to take or give their right of way based on social and physical context, \eg courteous behavior and low vehicular travel speed~\cite{hamilton2005improving}.
Compared to conventional traffic designs, shared spaces promote a more pleasant urban environment with slower motorized movement and safer and smoother traffic and less congestion~\cite{clarke2006evolution}.  
These designs nowadays can be found in urban areas of many European cities, like the Laweiplein intersection in the Dutch town Drachten, Skvallertorget in Norrk{\"o}ping, Kensington High Street in London~\cite{hamilton2008shared}, and the shared space in the German town Bohmte~\cite{bode2009verkehrsuntersuchung}, as well as in many other countries\footnote{\url{https://en.wikipedia.org/wiki/Shared\_space}}.

\begin{figure}[tb]
\centering
\vspace{2mm}
\includegraphics[trim=0in 3.5in 0in 0in, clip=true, width=1\linewidth]{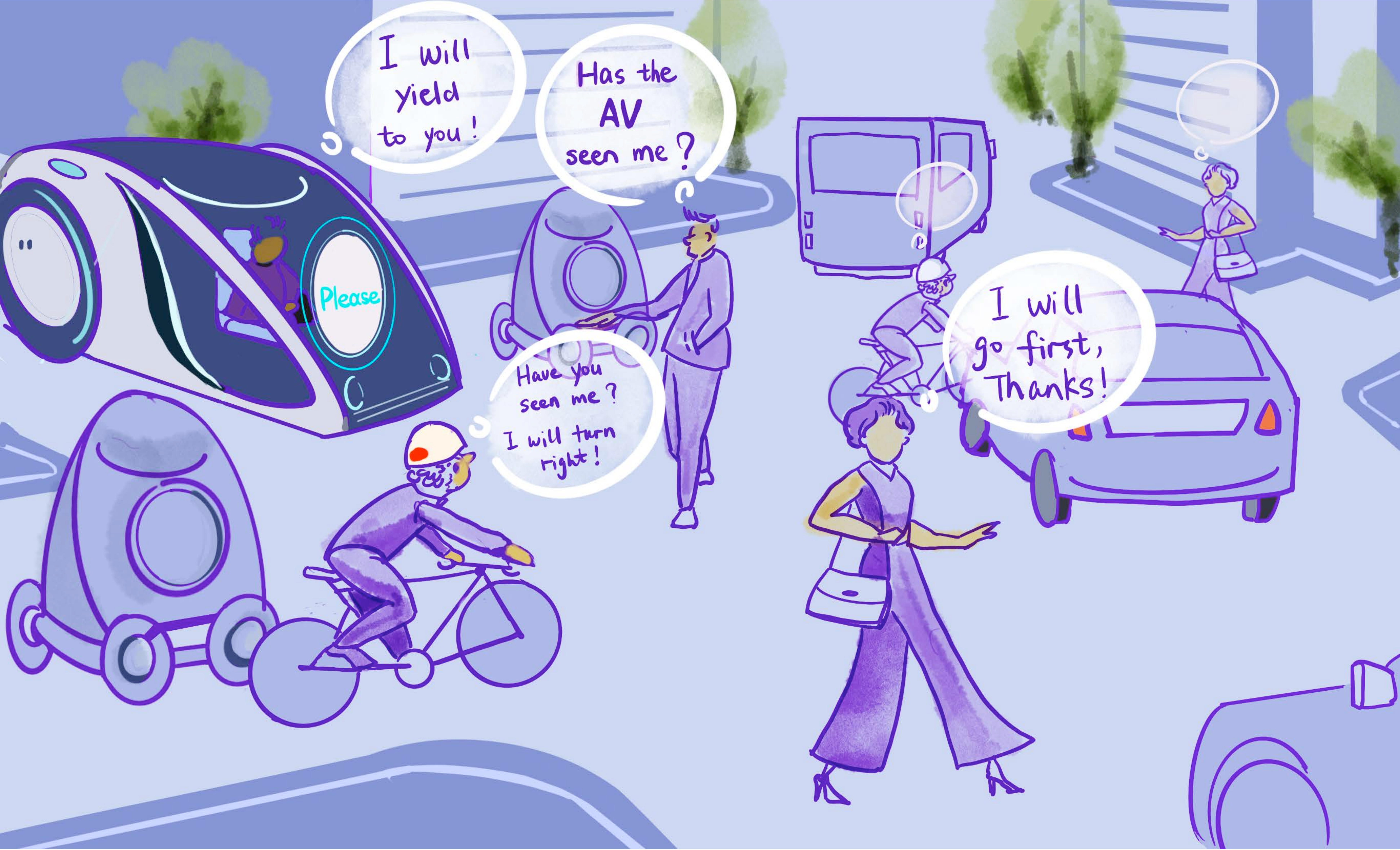}
\vspace{-5mm}
\caption{AV-VRU communication in a shared space.}
 \label{fig:Figure1}
\vspace{-6mm}
\end{figure}

However, with the advent of autonomous vehicles~(AVs), the established shared spaces may face new challenges with respect to traffic safety and smoothness, and frequent communication between AVs and vulnerable road users (VRUs) is needed (see Fig.~\ref{fig:Figure1}).
In this paper, we first review the challenges of the current shared-space designs and seek the potential solutions by leveraging eHMI to build the communication between AVs and VRUs, in order to seamlessly integrate AVs into shared-space traffic.

\subsection{Shared space as a traffic design}
In the past decades, various aspects have been considered in shared-space designs.
Sch{\"o}nauer~\cite{schonauer2017microscopic} reviewed the most important aspects, which are traffic flows, safety and accident aspects, driving speeds, parking demand, change in traffic behavior, and impact on urban development and land use.
In urban areas, shared spaces have been shown to improve both vehicle and pedestrian travel times and maintain safe and mixed use of road surface~\cite{wargo2016shared}.
Clarke~\etal~\cite{clarke2006evolution} found that the removal of traffic rules increases the perceived risk by vehicle drivers and therefore leads to a more cautious driving behavior, \eg driving at decreased speeds and more frequently giving way to VRUs; The shared-space designs have changed road users' behaviors, \eg cyclists start to use hand signals and road users seek eye contact with each other for communication.
In addition, shared spaces have been found to increase the functionality of public places, \eg the decrease of vehicle dominance increases pedestrian activities and dwelling time~\cite{karndacharuk2013analysis}.
 
Besides the positive aspects, shared spaces still remain controversial. 
On the one hand, in comparison to conventional traffic roads, they are designed to improve the activities of pedestrians and to achieve better traveler comfort and safety, where a slower vehicle speed is required and both drivers and pedestrians remain vigilant. 
On the other hand, people who oppose this concept argue that shared spaces may also lead to more pedestrian-vehicle conflicts \cite{kaparias2013analysis}.
One of the biggest concerns is about the safety of the elderly, children and other disabled road users, who have a limited capacity to perceive the surrounding environment~\cite{deichman2008shared}.
As a driver, the presence of the elderly and children is a very important factor regarding the driving behavior and the driver is supposed to pay more attention to them~\cite{kaparias2012analysing}. 
Another key issue is how to deal with high-volume road space.
At low vehicular traffic conditions pedestrians feel more comfortable~\cite{kaparias2012analysing}, whereas high volume road space increases the number of collisions between pedestrians and bicycles in non-motorized shared spaces~\cite{gkekas2020perceived}.
In addition, intention misunderstanding between drivers and pedestrians is the frequently stated problem in shared spaces~\cite{hamilton2005improving}. Pedestrians hope to get relevant information clearly by setting safe zones, where rich physical and multi-sensory cues in terms of surface tactility, color contrast and the enhancement of sound are provided \cite{parkin2012accounting}.

\subsection{Issues of AV-VRU interacting in shared spaces}
\label{subsec:AV-VRU-interacting}

In the foreseeable future, shared spaces will be populated with an unprecedented mixture of AVs and VRUs. Figure~\ref{fig:Figure1} depicts the mixed traffic in a shared space.
The advent of AVs will make the interactions even more complicated, where the shared-space designs purposely introduce ambiguities to hope for more cautious behavior~\cite{clarke2006evolution}.
When an unmanned AV, \ie~SAE level 3--5 AV~\cite{SAE_j3016b_2018}, and VRUs encounter in a shared space, the following issues with respect to their interactions should be considered:
\begin{itemize}
    \item[1)] The VRUs may have difficulties in perceiving or understanding the intentions of the AV because some of the conventionally used communication methods from the driver, \eg eye contact, head nod and hand gesture, will be altered or vanish~\cite{vissers2017safe, merat2018externally,liu2021importance}. 
    \item[2)] The low-speed driving makes it difficult for VRUs to obtain the intention and predict the behavior of the AV from its movement dynamics, especially acceleration and deceleration~\cite{Naoto2019}.
    \item[3)] The VRUs may feel not safe if they cannot easily understand the intentions of the AV~\cite{hamilton2005improving,kaparias2012analysing,liu2020gaze}. It is also likely to increase conflicts between them~\cite{merat2018externally}.
    \item[4)] The VRUs may become hesitant and nervous because of the distrust of the AV due to,~\eg the lack of knowledge of the AV's inner workings or logic~\cite{hoff2015trust,liu2021importance}. 
 \end{itemize}

The above issues challenge the justifications of the shared-space designs.
The potential conflicts between AVs and VRUs could reduce the safety and efficiency of the traffic flow~\cite{merat2018externally}, and the lack of commonly used explicit communication could reduce the traffic climate, leading to the so-called bad pro-sociality~\cite{sadeghian2020exploration} and hindering the public acceptance and broad deployment of AVs in shared spaces. 
Hence, in this paper we explore plausible solutions to these new challenges the shared-space traffic confronts with no or minimum alterations of the already established shared spaces in many places.  

\subsection{eHMI design as a solution to AV in shared spaces}
As mentioned above, AV-VRU interacting in shared spaces has several integrated and typical issues due to, \eg disappeared explicit communication from the driver, low-speed driving, misunderstanding AV's intention, mixed driving environment and ambiguous agreement, which needs frequent communication between AV and VRUs.
A novel explicit communication such as external human-machine interface (eHMI) could be one solution to these issues~\cite{schieben2018designing,Busch2018,uttley2020road,liu2021importance}. 
eHMI is an interface located on or projecting from the external surface of a vehicle that can convey information about its driving intention to the surrounding VRUs substituting the driver~\cite{info11020061,tabone2021vulnerable}.
eHMIs are especially helpful in low-speed scenario~\cite{Naoto2019}, \eg in a shared space, because they not only can be designed to express the current and future driving intentions to the VRUs and help them make quick decisions in ambiguous driving scenarios~\cite{dietrich2019projection,liu2021importance}, but also can improve the perceived safety of VRUs ~\cite{de2019external,liu2020gaze}.
Moreover, the communication cues of the eHMIs are beneficial for pro-social aspects~\cite{sadeghian2020exploration} because they are useful in improving trust among VRUs, meanwhile, giving them a feeling that the AV is polite, building a good traffic climate~\cite{tabone2021vulnerable}.

Some traditional eHMI signaling methods, such as blinkers, brake lights and headlights, are legally required and highly standardized~\cite{info11020061}.
These traditional methods may not be suitable for shared spaces and they should be tailored according to the AV-VRU interactions with the consideration of the uniqueness of shared space traffic
(see Sec.~\ref{subsec:AV-VRU-interacting}). 
A typical example is that, it can be difficult for an AV to use the traditional signaling methods to make VRUs understand its driving intentions and suggestions when VRUs and the AV have to negotiate the right of way.

Currently, some novel eHMI designs have been proposed~\cite{deb2019comparison}, such as display on vehicle, projection on road, light strip or light spot, as well as anthropomorphic eHMIs, amongst others~\cite{fridman2017walk}.
In addition, other information can be communicated, \eg by projecting virtual paths.
This can be the path that the AV is following and thus showing its intention explicitly, whereas an AV can also project paths relating to the VRUs: this can be the projected path which the AV assumes that an VRU is going to take.
It can, however, also be a ``safe path'' to allow both road users to safely interact.
Such a projection modality has been realized from a technical point of view in~\cite{Busch2018} by projecting road markings and signals to visually indicate an AV's intended behavior.

However, how to design effective eHMIs for share spaces is still not fully explored.
First and foremost, we need to understand what are the most important factors that guide the shared-space traffic and how can we adopt these factors for the eHMI designs? 
To be more specific, the research questions of eHMI designs in shared spaces are as follows:
\begin{itemize}
    \item[1)] 
    What information and the communication methods are expected by VRUs when the AV needs to communicate to them in shared spaces?
    \item[2)] Do VRUs expect that the AV communicates to them by only using traditional signaling methods such as turn signal, headlight, break light and horn?  
    \item[3)] Do VRUs need different eHMI communication methods with respect to their transport modes? 
\end{itemize}

\section{Experiment by an Online Survey}

To answer what information an AV needs to convey to VRUs, \ie~pedestrians and cyclists, the communications between VRUs and the driver in a manually driving vehicle when they interact in shared spaces were surveyed via an online questionnaire. 
We use this surveyed data to analyze their communications, and then to guide the eHMIs designs for AV-VRU communication.
The questionnaire includes the following questions corresponding to three transport modes:
\begin{itemize}
\item[Q1:] \textbf{As a driver} driving a car in a shared space, how do you intend to convey information to pedestrians and cyclists in real life?
\item[Q2:] \textbf{As a pedestrian} walking in a shared space, how do you expect the driver to convey information to you?
\item[Q3:] \textbf{As a cyclist} cycling in a shared space, how do you expect the driver to convey information to you? 
\end{itemize}
For each question, nine items can be multiple selected that are 1) headlight flashing, 2) turn signal, 3) brake light, 4) horn, 5) car movement, 6) hand gesture, 7) eye contact, 8) head nod, and 9) no communication.
The items 1) to 8) are all explicit communication methods except for 5) car movement.
The 9) no communication represents their willingness to communicate.
Meanwhile, headlight flashing, turn signal, brake light and horn are explicit communications via the vehicle's traditional signaling methods; hand gesture, eye contact and head nod are explicit communication methods via the driver's body behaviors.
After that, participants were asked to explain what specific information they wanted to convey regarding each selected communication method.

Thirty-one participants (20 males and 11 females) within the age range of 24\,--\,59 (mean: 32.2, standard deviation: 5.0) were invited to the questionnaire. 
All participants are from Europe, especially Germany and most of the them are familiar with the concept of shared space.
We used the publicly available video\footnote{\url{https://www.youtube.com/watch?v=qgYzyGvMqjo}} recorded in the shared space Sonnenfelsplatz Graz~\cite{schonauer2017microscopic} as an example to demonstrate the shared-space traffic and help them recall their experience in shared space.
Note that we discarded the data from the four participants without driving license for Q1 and the data from the two non-cycling participants for Q3.

\begin{figure}[tb]
\centering
\subfigure[Drivers to convey information to pedestrians and cyclists]{
\includegraphics[trim=0in 0.01in 0in 0in, clip=true, width=1\linewidth,height=3cm]{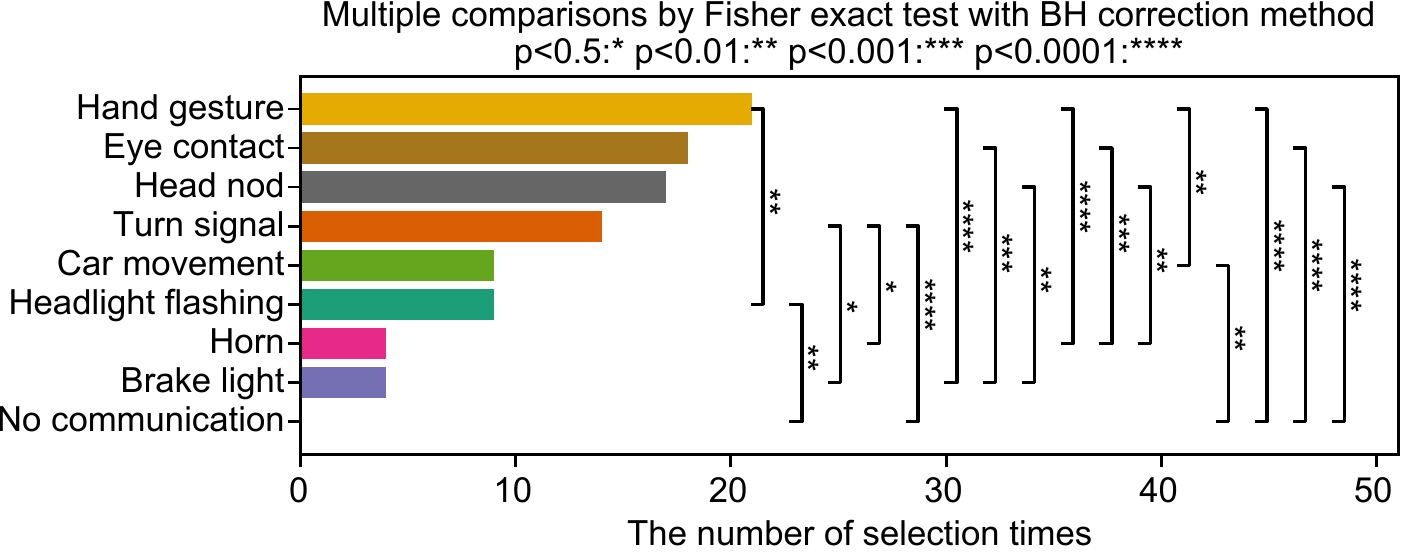}
\label{fig:driver}
}
\centering
\subfigure[Expected communication methods by pedestrians from drivers]{
\includegraphics[trim=0in 0.01in 0in 0.3in, clip=true, width=1\linewidth, height=2.6cm]{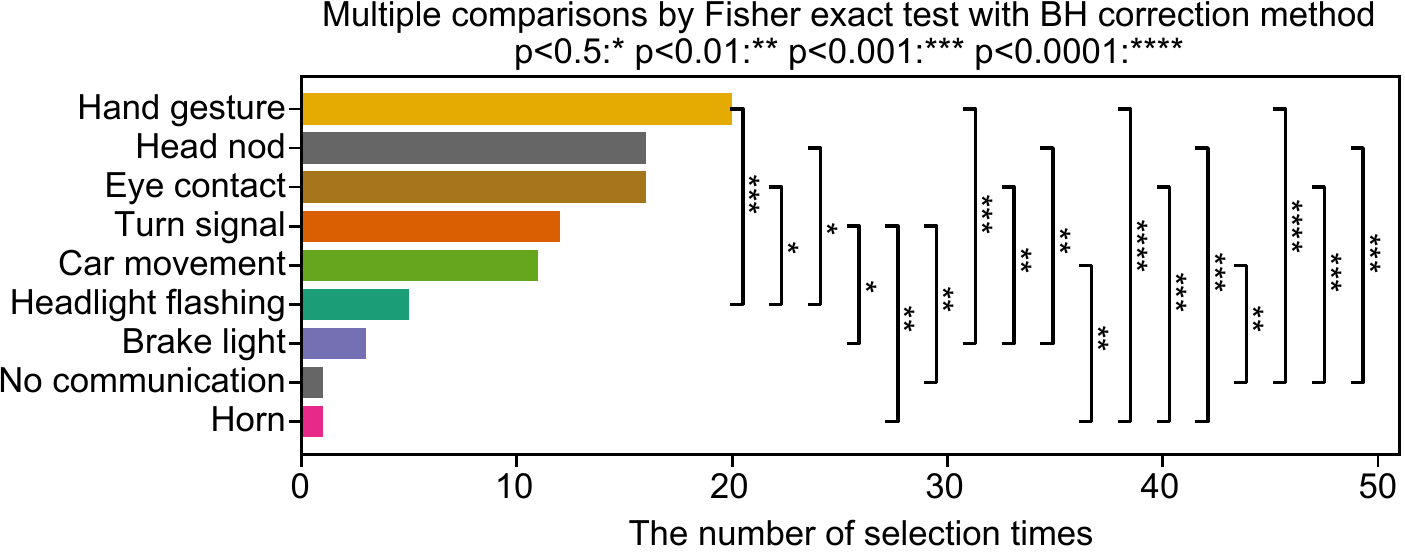}
 \label{fig:pedestrian}
}
\centering
\subfigure[Expected communication methods by cyclists from drivers]{
\includegraphics[trim=0in 0.01in 0in 0.3in, clip=true, width=1\linewidth, height=2.6cm]{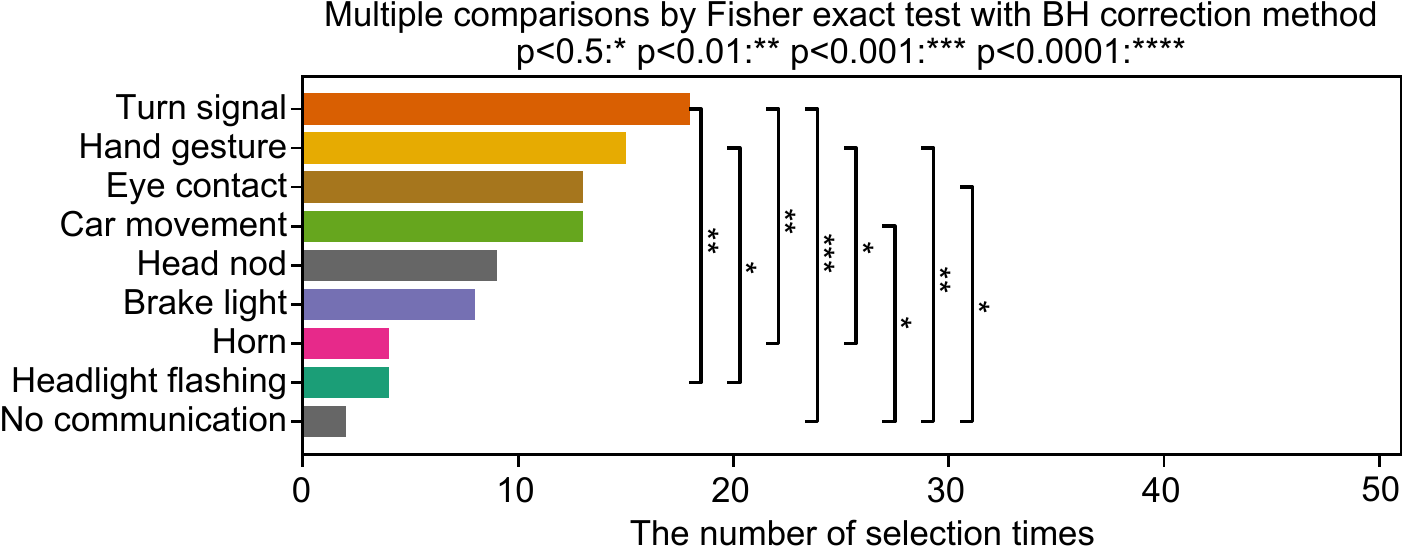}
 \label{fig:cyclist}
}
\vspace{-2mm}
\caption{Communication methods w.\,r.\,t transport modes.}
\vspace{-6mm}
\label{fig:interactionmethods}
\end{figure}

\begin{figure*}[tb]
\subfigure[Category~A]{
\includegraphics[width=0.333\linewidth]{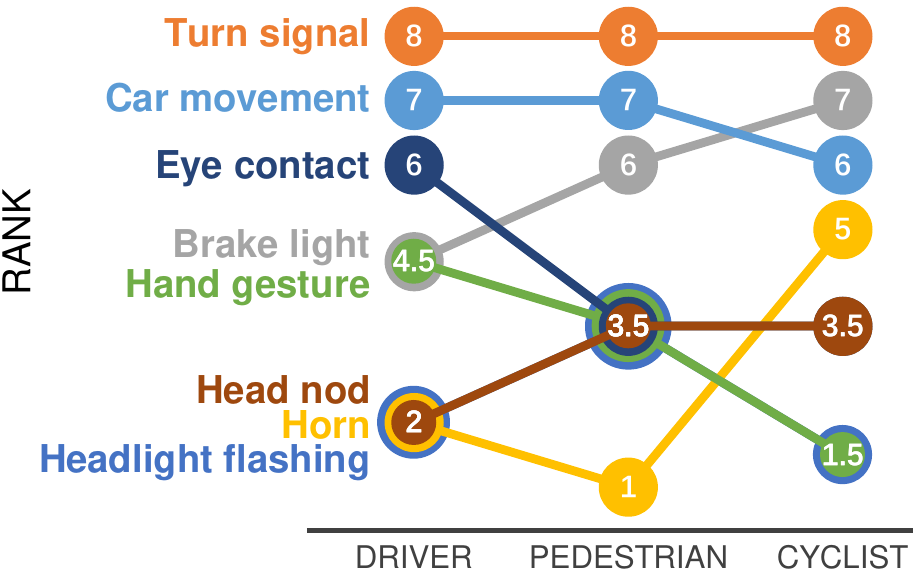}
 \label{fig:Category_A_rank}
}
\hspace{-4mm}
\subfigure[Category~B]{
\includegraphics[width=0.333\linewidth]{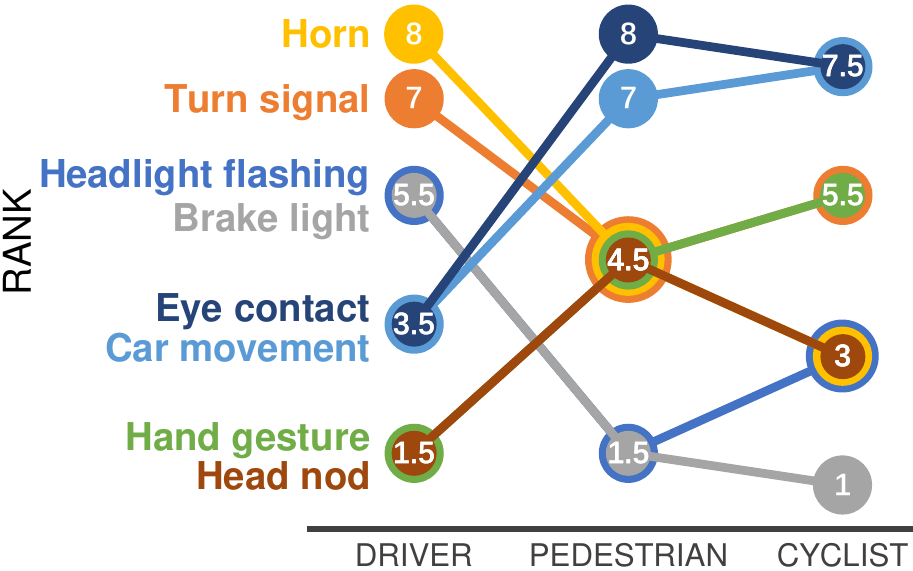}
 \label{fig:Category_B_rank}
}
\hspace{-4mm}
\subfigure[Category~C]{
\includegraphics[width=0.333\linewidth]{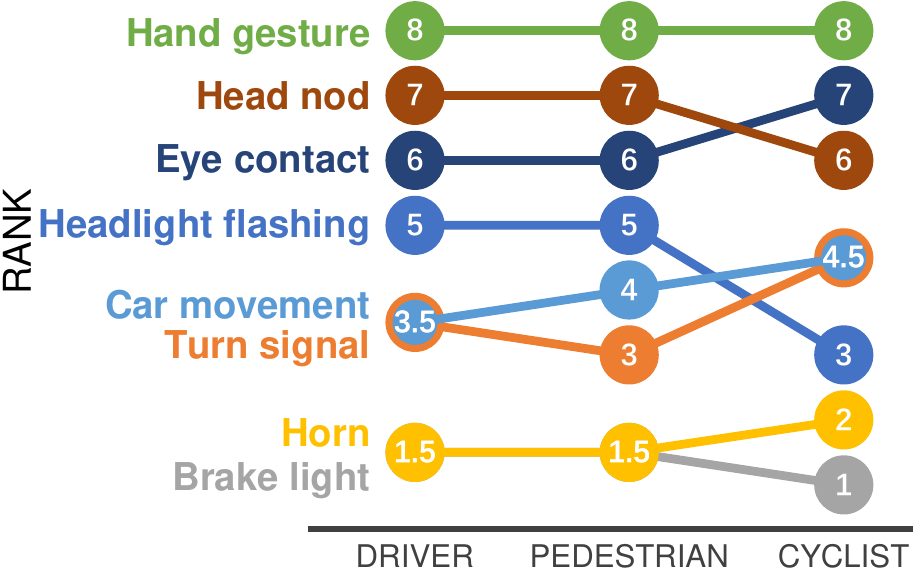}
 \label{fig:Category_C_rank}
 }
\vspace{-2mm}
\caption{The intended/expected communication method ranks for conveying information in the categories A, B and C.}
 \label{fig:Category_ranks}
\vspace{-5mm}
\end{figure*}

\section{Results}

\subsection{Communication methods  w.\,r.\,t. transport modes}

Here, we analyze the communication methods in a shared space intended by drivers and expected by pedestrians and cyclists from the drivers. The differences among multiple methods are compared by the Fisher exact test with the Benjamini-Hochberg (BH) correction.
The corresponding results are presented in Fig.~\ref{fig:interactionmethods} for each transport mode. The horizontal axis is the selection count and the vertical axis is the communication methods ranked by the counts.

Figures~\ref{fig:driver} and \ref{fig:pedestrian} illustrate that in a shared space, the communication methods intended by drivers and expected by pedestrians are highly aligned with each other.
The top three intended/expected communication methods were hand gesture, head nod and eye contact, which were significantly more frequently selected than implicit communication by car movement and explicit communication on vehicle by horn and brake light. 
However, in Fig.~\ref{fig:cyclist}, the expected communication methods from cyclists to receive information was not aligned with what drivers intended to convey (see Fig.~\ref{fig:driver}). 
Turn signal and car movement were expected by cyclists, which were not highly intended by drivers and pedestrians. 
Compared to pedestrians walking at low speed, the higher speed and less flexibility (\eg stabilizing the bicycle) of cyclists may impact their choice of methods to receive information from vehicles. The direct communication cue of turn signal is more straightforward to be understood by cyclists.
Hand gestures and eye contact are still in the top three communication methods selected, which are significantly more frequent than horn and headlight flashing.

\subsection{Communication methods to different information}
\label{subsec:categoty}

The communication method is a carrier of the information to be disseminated.
For information that expresses the same intent, road users may intend to use different methods associated with their transport modes, \ie~the methods drivers intend to use and the ones VRUs expect the drivers to use could be different.
To investigate this difference, we further categorized the information sent by the drivers to the VRU receivers (pedestrians and cyclists) into \textit{situation awareness}, \textit{risk evaluation} and \textit{decision making}, based on the criteria reported by Liu~\etal \cite{liu2021importance}.
In this way, we can analyze the effectiveness of an eHMI in terms of each category for the AV in shared spaces.

\textbf{Category~A}: information to help VRUs to be aware of situations. A driver conveys information about her/his intentions and the states of the car~\eg``I'm stopping'', ``I want to turn left/right'' and ``I will go''.

\textbf{Category~B}: information to help VRUs perceive risks. A driver conveys information about caution, such as ``danger'', ``warning'' and ``emergency situation''.

\textbf{Category~C}: information to help VRUs make decisions. This information includes suggestions and gratitude from drivers, such as ``you go first'', ``go ahead'' and ``thanks''.

Note that the information about gratitude from a driver, \eg ``thanks'', could be considered as positive feedback for the decision made by the VRUs.
It could encourage the VRUs to quickly make a decision that satisfies both parties in the next interaction with the driver.

For each category, the communication methods intended by drivers and expected by VRUs, denoted as \textit{driver-vs-pedestrian}, \textit{driver-vs-cyclist} and \textit{pedestrian-vs-cyclist} in Table~\ref{table:SCC_ABC}, were ranked by the number of selections. The corresponding ranks in each category are visualized in Fig.~\ref{fig:Category_ranks}, where the value in each circle denotes the selection number and different communication methods are color coded, and a large slope of the connected edges denotes a bigger rank difference among different transport modes.

The Spearman's correlation coefficient ($r_s$) was used to compare the ranks and the t-test to estimate the corresponding statistical significance.
The null hypothesis $H_0$ is that there is no correlation between the methods intended and expected in each category, \ie $r_s$ is $0$;
Only if $r_s$ close to $1$ and $p<0.05$, then there is a significantly strong correlation between them.
Otherwise, the categorized tendency and expectation are different (\ie not correlated) with respect to their transport modes.

\begin{table}[bt]
\vspace{2mm}
\small
\centering
\caption{Spearman's correlation among the communication method ranks for each information category}
\setlength{\tabcolsep}{0.5mm}{
\begin{tabular}{l|cc|cc|cc}
\hline
 & \multicolumn{2}{c|}{\textbf{Category~A}} & \multicolumn{2}{c|}{\textbf{Category~B}} & \multicolumn{2}{c}{\textbf{Category~C}} \\ \cline{2-7} 
\textbf{} &  $r_{s}$ & $p$-value & $r_{s}$ & $p$-value & $r_{s}$ & $p$-value \\ \hline
\textbf{driver-vs-pedestrian} & 0.804 & 0.016 & -0.312 & 0.452 & 0.994 & \footnotesize{\textless{}0.0001} \\ \hline
\textbf{driver-vs-cyclist} & 0.578 & 0.134 & -0.308 & 0.458 & 0.897 & 0.003 \\ \hline
\textbf{pedestrian-vs-cyclist} & 0.668 & 0.070 & 0.875 & 0.004 & 0.892 & 0.003 \\ \hline
\end{tabular}
}
\label{table:SCC_ABC}
\vspace{-6mm}
\end{table}

\begin{figure*}[t!]
\centering
\includegraphics[width=1\linewidth]{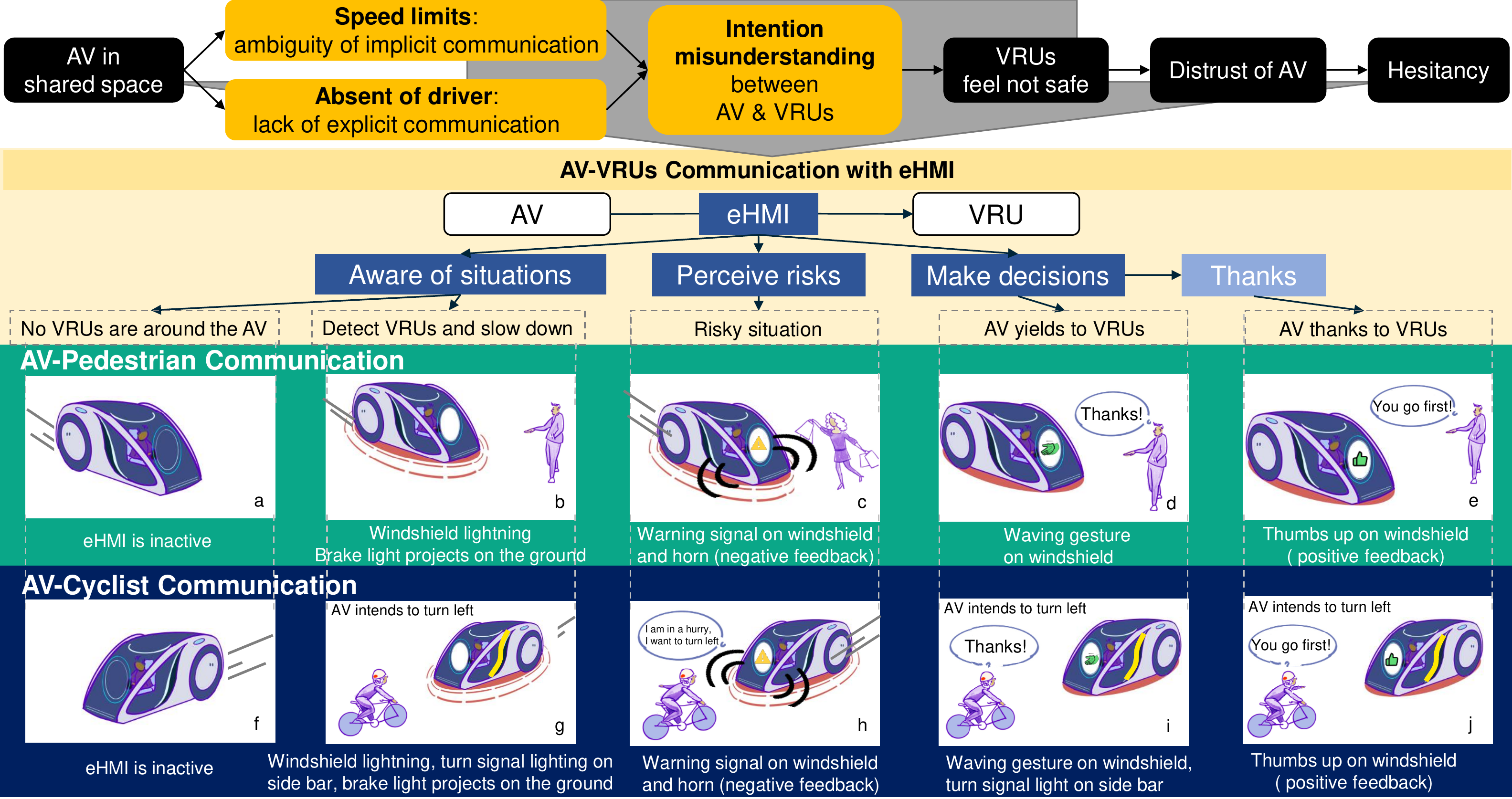}
\caption{The addressed issues and potential eHMI design concept.}
\vspace{-7mm}
\label{fig:eHMIs-prototype}
\end{figure*}

In the category~A, with respect to intention information, there was a significantly strong correlation ($r_{s}=0.804$, $p=0.016$) for driver-vs-pedestrian.
However, the correlation coefficients reduced to $r_{s}=0.578$ and $r_{s}=0.668$ for driver-vs-cyclist and pedestrian-vs-cyclist, and no significance was found. Moreover, Fig.~\ref{fig:Category_A_rank} shows the following patterns:
1) As a driver, turn signal and car movement were intended to convey her/his intentions, \eg ``I am turning to a certain direction'', ``I will drive now'' and ``I will stop now''.
Eye contact, hand gesture and brake light were also used to convey information in a smaller amount.
But head nod, horn and headlight flashing were not often used to convey the intention of the driver to help the VRUs to be aware of situations.
2) As a pedestrian, turn signal and car movement were expected to convey the information from the driver, which are aligned with what the driver was intended. 
Interestingly, pedestrians preferred the driver to use brake light over eye contact to convey her/his intentions.
3) As a cyclist, the ranks of turn signal, car movement and eye contact were in the upper position.
Meanwhile, horn was also required to convey information from the driver.
This indicates that compared to pedestrians, cyclists may expect information via clearer and more striking communication methods from the driver to help them obtain situation awareness.

In the category~B, pedestrians and cyclists expected communication methods regarding caution information from drivers are different from the ones the drivers intended to use.
There were non-significant correlations for driver-vs-pedestrian ($r_{s}=-0.312$, $p=0.452$) and driver-vs-cyclist ($r_{s}=-0.308$, $p=0.458$). 
In addition, the results also show that pedestrians and cyclists had a similar expectation for the communication methods from drivers, indicated by a strong pedestrian-vs-cyclist correlation ($r_{s}=0.875$, $p=0.004$).
Multiple intuitive examples can be seen in Fig~\ref{fig:Category_B_rank}.
Drivers intended to use horn, turn signal, headlight flashing and brake light to raise the caution of pedestrians and cyclists, whereas pedestrians and cyclists expected drivers to warn them through eye contact and car movement.
Besides, cyclists also expected that drivers could use obvious communication methods, \eg turn signal and hand gestures, to help them perceive risks correctly. These results are in line with the on-site study in a shared space setting and the questionnaires carried out by Merat~\etal \cite{merat2018externally}, which show that the information about an AV's actions, such as turning, stopping and acknowledgment of the detection of other road users was highly expected by the VRUs it encountered.

In the category~C, significant positive correlations ($r_{s}>0.890$, $p<0.01$) for driver-vs-pedestrian, driver-vs-cyclist and pedestrian-vs-cyclist were found for making decisions.
Figure~\ref{fig:Category_C_rank} also shows that the communication method ranks intended/expected by them were almost the same.
The top three of intended/expected communication methods were all conveyed through the driver's body behaviors, \ie hand gesture, head nod and eye contact.
These results are consistent with the study reporting that active communications through the body were often used in shared spaces, especially for cyclist~\cite{clarke2006evolution}.

In summary, participants thought that the communication via the driver's body behaviors was necessary when the driver wanted to give some suggestions or indicate intentions to the VRUs in a negotiation. 
Besides, when the driver conveyed information about her/his intentions and cautions to the VRUs, participants selected different communication methods with respect to their transport modes. 
These results suggested that novel eHMIs might be useful for AV-VRU communication when the original drivers are not present, and the eHMIs need to be dedicated to VRUs' transport modes to meet their distinctive expectations. 

\section{Potential eHMI Design Concept}
Although, the questionnaire mainly looked at mimicking (human) communication, the availability of eHMIs will also allow for new methods, such as windshield display, light bar around the AV and the brake light projects on the ground. 
Furthermore, the eHMI design concept could be used for improving VRUs' feeling of safety, reducing distrust and hesitancy~\cite{liu2021importance}.
Most importantly, the explicit information shown on eHMIs should match the implicit information~\cite{Dey2020}, \eg AV movement, otherwise, it may result in distrust between VRUs and AV. 

We propose a framework with a potential eHMI design concept to address the aforementioned issues of AVs driving in shared spaces. In the design component for AV-VRU communication, the eHMI designs respond to, \ie~situation awareness, risk perception, decision making, and gratitude based on the analysis of the replies of the questionnaire. As a prototype, some scenarios are used to demonstrate the functionalities of the eHMIs, more specifications of design will be explored in our future work. 

First, fig.~\ref{fig:eHMIs-prototype}~(a) and (f) are used as baselines with no eHMI displayed, if there is no VRU in the AV's neighborhood. 

In response to intention misunderstanding, the eHMI lightning on windshield and the brake light projecting on the ground are used to reduce the ambiguity of implicit communication due to speed limit.
It aims to raise the VRUs' awareness of the situation, \eg informing the VRUs that the AV has recognized them and currently is decelerating.

We located the eHMI on windshield at driver's place to meet the expected communication methods by VRUs from the driver's body, such as hand gesture. 
In this case, a brake light pulsing can even present a strong sense that the AV is still moving with deceleration but has not yet fully stopped (Fig.~\ref{fig:eHMIs-prototype}~(b) and (g)).
Note that this eHMI design is not targeted on any particular VRU, but only explicitly showing the AV's intention to the VRUs in the vicinity.

In a potential danger, apart from horn signal,
an extra visualization warning signal will be displayed on the windshield to warn the running pedestrian (Fig.\ref{fig:eHMIs-prototype}(c)) and the offensive cyclist (Fig.\ref{fig:eHMIs-prototype}~(h)) of the risk of collision, so that the VRUs can notice the AV promptly. 

Hand gestures are used to reduce the ambiguity and give positive feedback in communications. For example, waving gesture can be used by the AV to give the right way to the VRUs (Fig.\ref{fig:eHMIs-prototype}(d) and (i)) after confirming that they can interact safely by in-vehicle cameras and sensors, and a thumb-up to express gratitude to the VRUs' courteous behavior (Fig.\ref{fig:eHMIs-prototype}~(e) and (j)). 

Moreover, side bar lighting of the AV, as a more direct and obvious turn signal, can be used to emphasize the car movement with the aim to help cyclist understand the AV's intention and make a quick decision (Fig.\ref{fig:eHMIs-prototype}~(g) and (i)).

\section{Discussion}

In this section, we further discuss the benefits of the eHMIs on not only the sociality in shared spaces but also the autonomous driving systems.

\subsection{Effect of eHMI on the sociality in shared spaces}

From the results of the questionnaire, we found that there is positive feedback between the drivers and the VRUs, such as saying ``thank you'' after the negotiation.
This positive feedback makes them understand that their previous decisions are received with praise and gratitude from each other.
Including gratitude design element (\eg Fig.\ref{fig:eHMIs-prototype}(e) and (j)) may strengthen mutual trust between them and improve sociality in shared spaces.
For the AV, if it can give this positive feedback via the eHMI when interacting with VRUs in shared spaces, then the sociality of the AV and the VRUs' trust in the AVs could be promoted~\cite{liu2021importance}.

On the other hand, we also consider that the negative feedback is important as well for AV-VRU interactions, such as warnings.
According to the risk homeostasis theory~\cite{wilde1982theory}, if the allowable level of risk for the VRUs is high, then they may tolerate the actual receipted risks and take some risky behaviors, especially when they over-trust themselves~\cite{liu2019overtrust}.
In this situation, if the AV conveys a warning message through eHMI, it may promote pedestrians to calibrate the endurable level of risk and their over-trust in themselves.

In summary, along with the driving intention information disseminated by eHMIs, appropriate positive and negative feedback from the AV is beneficial for calibrating the trust of VRUs in the AVs and themselves. 
A harmonious and mutually trustworthy traffic society could be possibly formed through an appropriate communication and feedback loop between AVs and VRUs in shared spaces.

\subsection{Effect of eHMI on the autonomous driving systems}
In recent years there has been a large body of research on interaction modeling among road users for autonomous driving in urban areas of mixed traffic, as well as in shared spaces~\cite{nagel1992cellular,helbing1995social,lee2017desire,sadeghian2018sophie,chandra2019traphic,park2020diverse,cheng2020exploring}.
The most well-known conventional approaches are, \eg rule-based approaches such as social force model~\cite{helbing1995social} and cellular autometa~\cite{nagel1992cellular}.
Game theory~\cite{myerson2013game} is also applied to mimic the negotiation among road agents in shared spaces to achieve the equilibrium for each single agent~\cite{cheng2020trajectory}.
Early machine learning approaches rely on manual extracted features to stimulate complex decision-making process in interactions, such as Gaussian processes~\cite{wang2007gaussian} and Markov decision processing~\cite{kitani2012activity}.
In recent years, deep learning~\cite{lecun2015deep} approaches are trained to automatically learn interactions from large amount of real-world data.
The most widely used approaches are recurrent neural networks with long short-term memories~\cite{alahi2016social,chandra2019traphic} and deep convolutional neural networks~\cite{cui2019multimodal}.
In addition, deep generative models are employed to learn the multi-modalities of interactive behaviors, such as generative adversarial nets~\cite{gupta2018social} and variational auto-encoder~\cite{kingma2014auto,lee2017desire}.
Furthermore, attention mechanisms~\cite{xu2015show,vaswani2017attention} are incorporated into these models for modeling complex sequential patterns~\cite{sadeghian2018car,park2020diverse,cheng2021amenet} and reinforcement learning are applied to teach agents to behave like human road users~\cite{lee2017desire,co2018self}.
However, most of the works simplify the interaction process of road users as road agents considering only their motion behavior, and the possibilities of communications and feedback among them are over simplified or neglected.

On the other hand, some of the machine learning methods, especially deep learning models, have tried to extract explicit information from the training data to improve the accuracy of prediction, such as body pose~\cite{quintero2014pedestrian,ghori2018learning,shinmura2018estimation}, head direction~\cite{hasan2018mx}, gesture (raising arms)~\cite{pool2019context}, eye contact~\cite{onkhar2020automatic}, environmental scene context~\cite{sadeghian2018sophie,cheng2020mcenet}.
The major drawback of the approaches above is that even though some of the explicit information is encoded, there is no way for the agents to communicate with and provide feedback to each other.
There is still a gap in realistic interaction modeling.
One reason is that at the current research phase these explicit communications are difficult to be encoded into these models given the high complexity of individual behaviors. 
To this end, suggested by many other works and the findings in this paper, eHMIs are a supplementary way to mimic the communications between the automation systems and human road users, in order to establish effective and unambiguous understanding.
Hence, it is meaningful to take eHMIs into consideration when designing the automation systems for interaction modeling.

\section{Conclusion} 
This paper discussed the issues of interaction between AV and VRUs in shared spaces, \ie an AV lacks a way to communicate the VRUs when they want to reach an agreement.
To find a solution to these issues, we focused on the communication methods between manual driving vehicles and VRUs that have been popularized in shared spaces. 
A questionnaire was used to ask drivers, pedestrians and cyclists about how they would like a driver of a manually driving vehicle to communicate with VRUs in a shared space.
From the results, we found that the communication via the body behaviors of the driver, \eg hand gesture, head nod and eye contact, were expected by the pedestrians and cyclists, and even intended by the driver.
These communication methods were especially important when the driver wanted to suggest VRUs to reach an agreement.
In other words, some traditional eHMIs, \eg headlight, horn and brake light, were not often used for communication in this situation.
Besides, when the driver wanted to convey some information to help the VRUs obtain situation awareness and perceive risk, the intended/expected communication methods of drivers, pedestrians and cyclists were different.
Therefore, we considered that the novel eHMIs are useful to communicate with the VRUs replacing the original drivers in shared spaces, and a low-fidelity eHMI prototype was proposed to meet their various expectations.

In the discussion, we looked forward to the potential effects of eHMIs on the intelligent transportation.
We expect that the AVs can promote the formation of a harmonious and mutual trust transportation society through communication with VRUs via the eHMIs.
Besides, we expect that the AVs can induce/suggest VRUs to quickly make a unified decision by using the eHMIs.
This may indirectly improve the accuracy of AV prediction of VRUs behavior and reduce the error range of the prediction.

Nevertheless, our current study is limited to only a relatively small number of participants and no on-site experiments with AVs were carried out. However, the results are highly in lines with other recent studies that also seek eHMIs to improve the traffic safety and smoothness in a shared-space setting~\cite{merat2018externally} for a trustworthy~\cite{liu2021importance} and socially acceptable traffic climate~\cite{sadeghian2020exploration,tabone2021vulnerable}. In addition, our study is dedicated to shared spaces in general, and we hope that this preliminary study can pave the road to more in-depth research on analyzing autonomous driving in shared spaces.  

In future, we would like to further propose design guidelines for designing eHMIs from the perspective of VRUs, and establish an evaluation metric for AV-VRU interaction in shared spaces.
To achieve this goal, the AV-VRU interaction model and the trust process of VRUs in AV will be discussed based on a cognitive-decision-behavior model of VRUs proposed in~\cite{liu2020timing,liu2021importance}.
Furthermore, we would like to realize the AV-VRU communication and the high-precision prediction of VRUs' behaviors based on the interaction model.
Also, the communication methods for interactions between the AVs and the drivers in the manually driving vehicles also should be designed.
Moreover, in the future work, communications between multiple AVs and multiple VRUs should be considered as well, not only the interaction between single AV and VRU.
The visualization of virtual information can also be realized via augmented reality, which also allows VRUs to communicate their virtual paths~\cite{kamalasanan2020behaviour}.
In this way, a virtual infrastructure can be created, which potentially allows users to fall back to established rules.
As a matter of fact, such (potentially highly dynamic) virtual infrastructures have to be investigated further in real world scenarios.

\section*{Acknowledgments}
This work was supported by JSPS KAKENHI Grant Number JP20K19846, Japan and the Deutsche Forschungsgemeinschaft (DFG, German Research Foundation) - 227198829/GRK1931.
Yang Li is supported by China Scholarship Council (CSC) (No.~201906260302) at Karlsruhe Institute of Technology (KIT), Germany. 

\bibliographystyle{ieeetr}
\bibliography{mybib_norepeat}

\end{document}